\documentclass[a4paper]{aa}
\usepackage{psfig,times}
\begin{document}
\title{Mixing zones in magnetized differentially rotating stars}
\author{V.~Urpin\inst{1,2}}
\offprints{V. Urpin}
\institute{$^{1)}$ Departament de Fisica Aplicada, Universitat d'Alacant,
           Ap. Correus 99, 03080 Alacant, Spain \\
           $^{2)}$ A.F.Ioffe Institute of Physics and Technology and
           Isaac Newton Institute of Chili, Branch in St. Petersburg,
           194021 St. Petersburg, Russia}

\date{\today}

\abstract{We study the secular instability of magnetized differentially
rotating radiative zones taking account of viscosity and magnetic and 
thermal diffusivities. The considered instability generalizes the 
well-known Goldreich-Schubert-Fricke instability for the case of a 
sufficiently strong nagnetic field. In magnetized stars, instability 
can lead to a formation of non-spherical unstable zones where weak 
turbulence mixes the material between the surface and interiors. 
Such unstable zones can manifest themselves by a non-spherical
distribution of abundance anormalies on the stellar surface.   
}
\authorrunning{V. Urpin}
\titlerunning{Mixing zones in magnetized stars}

\maketitle

\section{Introduction}

The effect of the magnetic field on stellar differential rotation 
was considered by many authors. In the pioneering work, Fricke (1969) 
first pointed out the importance of thermal diffusion for stability 
of magnetized differentially rotating stars. This conclusion had a 
futher development in the paper by Acheson (1978) who took into 
account consistently the main diffusive processes (viscosity, thermal 
and magnetic diffusion) and noted that even such a weak kinetic 
effect as viscosity can influence the secular stability of magnetized 
stellar radiative zones. Acheson (1978) reviewed in detail the so 
called double diffusive instabilities that can occur in stellar 
radiative zones because of dissipative effects. Perhaps the best 
known in astrophysical context diffusive instability is that 
considered by Goldreich \& Schubert (1967) and Fricke (1968). In 
non-magnetic stars, this instability arises if the angular velocity 
depends on the height above the equatorial plane and can be 
responsible, for example. for the mixing and angular momentum 
transport in stellar radiative zones (see, e.g., Knobloch \& Spruit 
1982, Korycansky 1991). Note that some double diffusive instabilities 
can be of importance for the rotational evolution of degenerate stars 
as well (Urpin 2003a, 2003b). 

Most likely, however, that the magnetic field may exist in stellar 
radiative zones despite many uncertainties regarding the origin,
evolution and geometry of these fields. It seems that even magnetic 
buoyancy cannot substantially reduce the field in radiative zones
because of a stabilizing influence of stellar rotation on magnetic 
buoyancy instability. Generally, rotation can suppress this 
instability if the rotation velocity, $V$, is larger than the 
Alfv\'{e}n velocity, $V > c_{A}$ (see, e.g., Gilman 1970, Acheson 
\& Gibson 1978, Acheson 1979).  

Secular instabilities are typically sensitive to the magnetic field,
and the stability properties of magnetic and non-magnetic secular 
modes can be substantially different. The influence of the toroidal 
magnetic field on the Goldreich-Schubert-Fricke instability has been 
considered by Acheson (1978) who noted that such a field plays a 
destabilizing role and can change qualitatively the criterion of 
rotational instability. This effect can be of importance, for example, 
for the sun where the toroidal field is expected to be buried deep 
inside the radiative zone (Schmitt \& Rosner 1983). Note that a 
destabilizing influence of the toroidal field is not surprising 
because such field leads to instability even in non-rotating stars 
(see Tayler 1973, Goossens \& Tayler 1980). 

The effect of the poloidal field can generally be more complicated
and, to the best of our knowledge, has not been considered in its 
full generality. Recently, Menou, Balbus \& Spruit (2004) considered 
diffusive instabilities that occur in magnetized differentially
rotating stellar radiative zones and obtained the criteria of  
instability in the limit when either viscosity or magnetic diffusivity 
is vanishing. The consideration, however, was restricted by the case 
of a very weak magnetic field. The authors have found that stability 
properties are sensitive to the combination of viscosity, magnetic 
and thermal diffusivities and, following Acheson (1978), concluded 
that all three should generally be included in the analysis from the 
very beginning. The results of this consideration have been applied by 
Menou (2004) to account for millisecond oscillations in differentially 
rotating atmospheres of accreting weakly-magnetized neutron stars. 

The present paper extends the study of secular instability to the 
case of a relatively strong magnetic field and arbitrary diffusivities. 
We show that such the magnetic field changes drastically the stability
properties of radiative zones. In a strong field, the instability
of differential rotation can lead to a formation of non-spherical 
unstable zones where weak turbulence is generated. The geometry of 
these zones depends on the magnetic configuration and rotation. 
For example, in the simplest case of an axisymmetric field the 
turbulent zones are formed near the rotation axis and around the 
equator. If the field is not axisymmetric then the instability will 
lead to the formation of cone-like unstable zones within the 
radiative region. Turbulent motions mix the material in the unstable 
zones, thus a fraction of the material from deep interiors may reach 
the surface where chemically pecular regions should be formed.      

An outline of the paper is as follows. In Section, we derive the 
general dispersion equation taking account of viscosity, magnetic and 
thermal diffusivities. The criteria of instability are discussed in 
Section 3, and the growth time of secular instability is considered in 
Section 4. A discussion of results follows in Section 5.

\section{Basic equations and the dispersion relation}

The MHD equations governing the velocity, magnetic field and thermal
balance in the stellar radiative zone are 
\begin{eqnarray}
\dot{\vec{v}} + (\vec{v} \cdot \nabla) \vec{v} = - \frac{\nabla
p}{\rho} + \vec{g} + \frac{1}{4 \pi \rho} (\nabla \times \vec{B})
\times \vec{B} + 
\nonumber \\
\nu \left[ \Delta \vec{v} + \frac{1}{3} \nabla
(\nabla \cdot \vec{v}) \right], 
\end{eqnarray}
\begin{equation}
\dot{\rho} + \nabla \cdot (\rho \vec{v}) = 0, 
\end{equation}
\begin{equation}
\dot{\vec{B}} - \nabla \times (\vec{v} \times \vec{B}) = -
\nabla \times [\eta \cdot ( \nabla \times 
\vec{B})], 
\end{equation}
\begin{equation}
\nabla \cdot \vec{B} = 0, 
\end{equation}
\begin{equation}
\rho c_{p} \dot{T} - \beta \dot{p} + \rho c_{p} \vec{v} \cdot 
(\Delta \nabla T) = \nabla \cdot (\kappa \cdot \nabla T)
\end{equation}
(see, e.g., Landau \& Lifshitz 1981). Our notation is as follows:
$\vec{v}$ is the fluid velocity; $\rho$, $p$, $T$ are the density,
pressure and temperature, respectively; $\vec{B}$ is the magnetic
field, $\vec{g}$ is gravity, $\Delta \nabla T = \nabla T - 
\nabla_{ad}T$ is a difference of the real and adiabatic temperature 
gradients, $\nabla_{ad} T = \beta \nabla p/ \rho c_{p}$; $\beta = - 
(\partial \ln \rho/\partial \ln T)_{p}$, $c_{p}$ is the specific 
heat at constant pressure; $\nu$ is the kinematic viscosity; 
$\eta$ and $\kappa$ are the magnetic diffusivity and thermal 
conductivity, respectively. We ignore the spatial dependence of 
the kinetic coefficients $\nu$, $\eta$, and $\kappa$, which is 
appropriate for a short wavelength analysis. 

We work in cylindrical coordinates ($s$. $\varphi$, $z$) with the 
unit vectors ($\vec{e}_{s}$, $\vec{e}_{\varphi}$, $\vec{e}_{z}$)
and assume that the fluid rotates with the angular velocity 
$\Omega=\Omega(s, z)$ in the unperturbed state. We assume the 
Alfv\'en speed, $c_{\rm A}=B/ \sqrt{4 \pi \rho}$, to be small compared 
to the sound speed, $c_{\rm s}$. Then, the magnetic field should satisfy 
the condition
\begin{equation}
B < 3 \times 10^{7} \sqrt{ \rho T_{6}} \;\;\; {\rm G}, 
\end{equation}
where $\rho$ is measured in g/cc and $T_{6} =T/10^{6}$ K. This 
assumption still leaves a wide range of $B$ which provide an appreciable 
influence on the stability properties and are of interest for stellar 
radiative zones. Note that under the condition $c_{\rm s} > c_{\rm A}$, 
the Alfv\'en speed can be larger than the rotation velocity $s \Omega$. 

In the unperturbed state, the star is assumed to be in hydrostatic 
equilibrium in the $s$- and $z$-directions,
\begin{equation}
\frac{\nabla p}{\rho} = \vec{G} \;\;, \;\;\;\;
\vec{G} = \vec{g} + \Omega^{2} \vec{s}
+ \frac{1}{4 \pi \rho} (\nabla \times \vec{B}) \times \vec{B}. 
\label{4}
\end{equation}
Generally, even if $c_{\rm s} > c_{\rm A}$ and the unperturbed Lorentz 
force is small compared to the pressure force, the effect of the magnetic
field cannot be neglected in this equation. Taking the curl of equation 
(7), we obtain
\begin{equation}
\frac{1}{\rho^{2}} \nabla p \times \nabla \rho = 
\vec{e}_{\varphi} s  \frac{\partial \Omega^{2}}{\partial z} +
\vec{L} \; , 
\end{equation}
where
$$
\vec{L} = \frac{1}{4 \pi} \nabla \times
\left[ \frac{1}{\rho} (\nabla \times \vec{B}) \times \vec{B}
\right]. 
$$
Therefore, the magnetic field (as well as rotation) can be responsible 
for the baroclinic effect that is oftenly crucial for stability. 
The Lorentz force splits generally into potentional (magnetic
pressure) and solenoidal (magnetic tension) parts. In our study, the 
effect of the both parts on the unperturbed state is taken into 
account consistently. 

Consider stability of axisymmetric short wavelength perturbations
with the spacetime dependence $\propto \exp ( \gamma t - i \vec{k}
\cdot \vec{r})$. where $\vec{k}= (k_{s}, 0, k_{z})$ is the wavevector,
$|\vec{k} \cdot \vec{r}| \gg 1$. Small perturbations will be indicated 
by subscript 1, whilst unperturbed quantities will have no subscript, 
except for indicating vector components. The linearized MHD-equations 
read in the local approximation
\begin{eqnarray}
(\gamma + \omega_{\nu}) \vec{v}_{1} + \frac{\nu}{3} \vec{k} (\vec{k}
\cdot \vec{v}_{1}) + 2 \vec{\Omega} \times \vec{v}_{1} 
+ s \vec{e}_{\varphi} (\vec{v}_{1} \cdot \nabla \Omega)   
\nonumber \\
= i \vec{k} \frac{p_{1}}{\rho} + \vec{G} 
\frac{\rho_{1}}{\rho} - \frac{i}{4 \pi \rho} (\vec{k} \times \vec{B}_{1})
\times \vec{B},  
\end{eqnarray}
\begin{equation}
\gamma \rho_{1} + \vec{k} \cdot \vec{v}_{1} = 0, 
\end{equation}
\begin{equation}
(\gamma + \omega_{\eta}) \vec{B}_{1} -
s \vec{e}_{\varphi} ( \vec{B}_{1} \cdot \nabla \Omega)  
= - i (\vec{k} \cdot \vec{B}) 
\vec{v}_{1}, 
\end{equation}
\begin{equation}
\vec{k} \cdot \vec{B}_{1} = 0, 
\end{equation}
\begin{equation}
(\gamma + \omega_{\chi}) T_{1} - \frac{\gamma \beta}{\rho c_{p}} p_{1}
= - \vec{v}_{1} \cdot (\Delta \nabla T),
\end{equation}
where the characteristic inverse timescales of the viscous and
ohmic dissipation and the thermal diffusion are given by $\omega_{\nu}= 
\nu k^{2}$, $\omega_{\eta} = \eta k^2$, and $\omega_{\chi}= 
\chi k^{2}$, respectively; $\chi = \kappa / \rho c_{p}$ is the thermal
diffusivity. 

Perturbations of the pressure in equation (9) can be 
expressed in terms of perturbations of the density and temperature, 
using the equation of state,
\begin{equation}
\frac{p_{1}}{p} = \left( \frac{\partial \ln p}{\partial \ln \rho} 
\right)_{T}
\left( \frac{\rho_{1}}{\rho} + \beta \frac{T_{1}}{T} \right).
\end{equation} 

We consider the dispersion relation in the case $kc_{\rm s} \gg \gamma$ 
that corresponds to the Boussinesq approximation for a non-magnetic
fluid. Apart from this, the applicability of the Boussinesq 
approximation in the magnetic field imposes a restriction on the 
field strength. Namely, the magnetic field should satisfy the 
inequality $c_{A}^2 /c_{s}^2 \ll (kH)^{-1}$ where $H$ is the 
lengthscale of unperturbed quantities. Solving equations (9)-(14) 
under these two conditions, we obtain the following dispersion relation
\begin{equation}
\gamma^{5} + a_{4} \gamma^{4} + a_{3} \gamma^{3} + a_{2} \gamma^{2}
+ a_{1} \gamma + a_{0} = 0 \; , 
\label{10}
\end{equation}
where
\begin{eqnarray}
\lefteqn{a_{4} = 2 \omega_{\eta} + 2 \omega_{\nu} + \omega_{\chi},}
\nonumber\\
\lefteqn{a_{3}= (\omega_{\eta} + \omega_{\nu})^{2} \!\!
+ 2 (\omega_{\eta} \omega_{\nu} \! + \!\omega_{\chi} \omega_{\eta} 
\! + \! \omega_{\nu} \omega_{\chi} \!) \! 
+ \! 2\omega_{A}^{2} \! + \! \omega_{g}^{2} \! + \! Q^{2} ,} 
\nonumber \\
\lefteqn{a_{2} = \omega_{\chi} (\omega_{\nu} + \omega_{\eta})^{2} +
2 \omega_{\eta} \omega_{\nu} (\omega_{\nu} + \omega_{\eta} +
\omega_{\chi}) + }
\nonumber \\
&& \quad \omega_{\nu} (\omega_{g}^{2} \! + \! 2 \omega_{A}^{2}) \! + \!
2 \omega_{\eta} (\omega_{g}^{2} \! + \! \omega_{A}^{2} \! + 
\! Q^{2}) \! + \!
\omega_{\chi}(Q^{2} + 2 \omega_{A}^{2}) ,
\nonumber \\
\lefteqn{a_{1} = 2 (\omega_{\nu} \omega_{\eta} + \omega_{A}^{2})
(\omega_{\nu} \omega_{\chi} + \omega_{\nu} \omega_{\eta} +
\omega_{\eta} \omega_{\chi}) + }
\nonumber \\
&& \quad \omega_{\eta} [ (\omega_{\eta} + 
2 \omega_{\nu}) \omega_{g}^{2} + (\omega_{\eta} + 2 \omega_{\chi}) Q^{2} 
- \omega_{\eta} \omega_{\nu}^{2}] +
\nonumber \\
&& \quad \omega_{A}^{2} \left( \omega_{g}^{2} +
\omega_{A}^{2} + F^{2} \right) ,
\nonumber \\ 
\lefteqn{a_{0} = \omega_{\eta}^{2} ( \omega_{\chi} \omega_{\nu}^{2}+
\omega_{\nu} \omega_{g}^{2} + \omega_{\chi} Q^{2}) + 2 \omega_{A}^{2}
\omega_{\nu} \omega_{\eta} \omega_{\chi} +}
\nonumber \\
&& \quad  \omega_{A}^{2} \left( \omega_{\eta} \omega_{g}^{2} + 
\omega_{\chi}
\omega_{A}^{2} + \omega_{\chi} F^{2} \right), 
\nonumber
\end{eqnarray}
and
\begin{eqnarray}
\lefteqn{Q^{2} = 4 \Omega^{2} \frac{k_{z}^{2}}{k^{2}} + F^{2} \;, \;\;
F^{2} = \frac{s k_{z}}{k^{2}}
\left( k_{z} \Omega_{s}^{2} - k_{s} \Omega_{z}^{2} \right) \;, }
\nonumber \\
\lefteqn{\Omega_{s}^{2} = \frac{\partial \Omega^{2}}{\partial s} \;, \;\;
\Omega_{z}^{2} = \frac{\partial \Omega^{2}}{\partial z} \;, \;\;
\omega_{A} = \frac{\vec{k} \cdot \vec{B}}{\sqrt{4 \pi \rho}} \;, }
\nonumber \\
\lefteqn{\omega_{\rm g}^{2} = - \vec{C} \cdot \left[ \vec{G} -
\frac{\vec{k}}{k^{2}} (\vec{k} \cdot \vec{G}) \right] \; , \;\;
\vec{C}= \frac{\beta}{T} \Delta \nabla T \;.} 
\nonumber
\end{eqnarray}
This equation describes five low-frequency modes that exist in 
differentially rotating and weakly magnetized non-ideal fluids. 
In the limit of ideal MHD ($\nu=\eta=\chi=0$), equation (15) 
reduces to a quadric one that describes four dynamical modes. 
The fifth root corresponds to a secular mode that appears 
only due to dissipative effects, and we address the properties 
of this mode. In a vanishing magnetic field ($B=0$, $\eta=0$), 
this mode has been considered by Goldreich \& Schubert (1967) 
and Fricke (1968). Equation (15) differs only by notations 
from the dispersion equations derived by Menou, Balbus \& Spruit 
(2004). Note, however, that these authors considered the stability
properties of Equation (15) only in a weak magnetic field, $\Omega 
\gg \omega_{A}$, and in either the inviscid or perfect-conductor
limit. In this paper, the secular stability will be considered
for a stronger magnetic field with $\omega_{A} > \Omega$ (but still
satisfying the condition $c_{s} > c_{A}$) and for arbitrary 
diffusivities

\section{Criteria of instability}

The Hurwitz theorem presents a sufficient condition of 
instability and states that equation (15) has unstable solutions
with a positive real part if one of the following inequalities
\begin{eqnarray}
\lefteqn{a_{4} < 0 \;, \;\;\; a_{0} <0 \;,}
\nonumber \\
\lefteqn{A_{1} \equiv a_{4} a_{3} - a_{2} < 0 \;, }
\nonumber \\
\lefteqn{A_{2} \equiv a_{2} (a_{4} a_{3} -a_{2}) - a_{4}(a_{4} a_{1} 
- a_{0}) <0 \;,}
\nonumber \\
\lefteqn{A_{3} \equiv (a_{4} a_{1} -a_{0}) [a_{2} (a_{4} a_{3} 
- a_{2}) - a_{4}
(a_{4} a_{1} - a_{0})] - }
\nonumber \\
&& \quad \quad \quad \quad \quad \quad - a_{0} (a_{4} a_{3} 
- a_{2})^{2} < 0 \;
\end{eqnarray}
is fulfilled (see Aleksandrov, Kolmogorov \& Laurentiev 1985). 
Since $\omega_{\nu}$, $\omega_{\eta}$, and $\omega_{\chi}$ are 
positive defined quantity, the condition $a_{4} < 0$ never applies. 

We consider only the criterion $a_{0} < 0$ that corresponds to
double diffusive instability and is a sufficient condition of this
instability. We focus on relatively large wavelengths, $\lambda = 
2 \pi / k$, for which $\omega_{A}^{2} \gg \omega_{\nu} \omega_{\eta}$. 
This condition can be rewritten as
\begin{equation}
\lambda > \lambda_{c} =2 \pi \frac{\sqrt{\nu \eta}}{c_{A}} = 2 \pi
\frac{\sqrt{4 \pi  \rho \nu \eta}}{B_{p}},  
\end{equation} 
where $B_{p}=\sqrt{B_{s}^{2} + B_{z}^{2}}$ is the strength of the
poloidal field. Mention that $\lambda$ should be small compared to 
$H$ because we use a local approximation. Following Menou, Balbus 
\& Spriut (2004), we can estimate $\rho \nu \eta \sim (2-6) \times
10^{3}$ g cm/s$^{2}$ in the radiative zone. Then, we have 
$\lambda_{c} \sim 10^{3}/B$ cm where $B$ is measured in Gauss. 
This leaves a comfortable range of wavelengths for which $H > 
\lambda > \lambda_{c}$ and our local analysis is valid. Note 
that although conditions for instability can indeed be found using
our approach with the accuracy in terms $\sim (\lambda_{c}/ 
\lambda)^2$, the critical wavelength that discriminates between 
stability and instability cannot be addressed by this analysis
because this wavelength is $\sim \lambda_{c}$.

Under the condition (17), the criterion $a_{0} < 0$ reads
\begin{equation}
\chi ( F^{2} + \omega_{A}^{2}) + \eta \omega_{g}^{2} < 0. 
\end{equation}
Equation (18) differs from the condition of instability
obtained by Fricke (1968) by the last term on the r.h.s.
Usually, $\eta/ \chi$ is small in stars but, on the contrary, the
ratio $\omega_{g}^{2}/ F^{2}$ is large. Therefore, whether or
not the contribution of the last term is small will depend crucially
on the particular system under consideration. This point concerning
viscosity and magnetic diffusivity has been first emphasized by 
Acheson (1978). The condition (18) also differs essentially from 
the well-known criterion of the magnetorotational instability,
\begin{equation}
F^{2} + \omega_{A}^{2} + \omega_{g}^{2} < 0 
\end{equation}
(e.g., Fricke 1969, Balbus \& Hawley 1991, Balbus 1995, Urpin 1996). 
Note that criterion
(19) is valid only if dissipative effects are negligible and the
magnetic field is very weak, $\omega_{A} \ll s |\Omega_{s,z}|$.
Compared to the magnetorotational instability, the stabilizing 
effect of stratification is much reduced in equation (18). Since 
usually $\omega_{g}^{2} \gg F^{2}$ in stars, the criterion (18) 
requires a much weaker differential rotation for instability than 
the criterion (19). 

In a realistic situation, all three terms in equation (18) 
should be taken into account. Generally, the buoyancy frequency  
is much larger than $\Omega$, and the stratification term is not 
negligible even in a fully ionized stellar plasma where $\eta$ is 
small. The inequality (18) depends on the direction 
of a wavevector and can be rewritten as follows
\begin{eqnarray}
N(\vec{k}) \equiv F^{2} + \omega_{A}^{2} + \frac{\eta}{\chi} 
\omega_{g}^{2} = 
\frac{\eta}{\chi} \omega_{0}^{2} + A \frac{k_{z}^{2}}{k^{2}} - 
\nonumber \\
D \frac{k_{s} k_{z}}{k^{2}} + 
E \frac{k_{s}^{2}}{k^{2}} < 0,
\end{eqnarray}
where
\begin{eqnarray}
&& A = s \Omega_{s}^{2} + \omega_{A0}^{2} \cos^{2} \alpha  + 
\frac{\eta}{\chi} C_{z} G_{z}, 
\nonumber \\
&& D = s \Omega_{z}^{2} - \omega_{A0}^{2} \sin 2 \alpha  -
\frac{\eta}{\chi} (C_{s} G_{z} + C_{z} G_{s}) , 
\nonumber \\
&& E = \omega_{A0}^{2} \sin^{2} \alpha + \frac{\eta}{\chi} C_{s} G_{s}. 
\nonumber 
\end{eqnarray}
In these expressions, we use notations
\begin{eqnarray}
\omega_{0}^{2} = - \vec{C} \cdot \vec{G} \;, \;\; 
\omega_{A0}^{2} = \frac{k^{2} B_{p}^{2}}{4 \pi \rho} \;, \nonumber
\end{eqnarray}
for the Brunt-V\"{a}is\"{a}l\"{a} and Alfven frequences, respectively;
$\alpha$ is the angle between the poloidal magnetic field and the axis 
of rotation. In what follows, we use the procedure proposed by Miralles, 
Pons \& Urpin (2004) to derive the criterion that does not depend on 
the direction of $\vec{k}$ and determines the minimal differential 
rotation leading to instability. Since the dependence of $N(\vec{k})$ 
on the direction of $\vec{k}$ is rather simple, we can obtain that 
minimum of $N(\vec{k})$ corresponds to
\begin{equation}
\frac{k_{z}^{2}}{k^{2}} = \frac{1}{2} 
\left[ 1 \pm \sqrt{\frac{(A-E)^{2}}{(A-E)^2 + D^2}} \right]~.
\label{phimax}
\end{equation}
Then, the minimum value of $N(\vec{k})$ yields the following condition 
of instability
\begin{equation}
s \Omega_{s}^{2} + \frac{\eta}{\chi} \omega_{0}^{2} + \omega_{A0}^{2} \pm
\sqrt{ D^{2} + (A-E)^{2}} < 0. 
\end{equation}
The gradients of $p$, $\rho$ and $T$ are related by
\begin{equation}
\frac{\nabla p}{p} = \left( \frac{\partial \ln p}{\partial \ln \rho} 
\right)_{T}
\left( \frac{\nabla \rho}{\rho} + \beta \frac{\nabla T}{T} \right).
\end{equation} 
It can be readily obtained from equation (8) that the condition of 
hydrostatic equilibrium leads to 
\begin{equation}
s \Omega_{z}^{2} = \left[\vec{C}\times\vec{G} \right]_{\varphi} =
C_z G_s -C_s G_z - L_{\varphi}.
\label{eqh2}
\end{equation}
Then, equation (22) can be further simplified to obtain
\begin{eqnarray}
s \Omega_{s}^{2} + \frac{\eta}{\chi} \omega_{0}^{2} + \omega_{A0}^{2} 
\pm \left\{ \left( s \Omega_{s}^{2} + \frac{\eta}{\chi} \omega_{0}^{2} 
+ \omega_{A0}^{2} \right)^{2} \! +  \right.
\nonumber \\
4 G_{z} \frac{\eta}{\chi} [ s(C_{z} \Omega_{s}^{2} \! - \! C_{s} 
\Omega_{z}^{2}) +
\omega_{A0}^{2} ( C_{z} \cos 2 \alpha  \! + \! C_{s} \sin 2 \alpha)]
\nonumber \\
- 4\sin^{2} \alpha \omega_{A0}^{2} \left( \frac{\eta}{\chi} 
\omega_{0}^{2} \! + \! s \Omega_{s}^{2} \right) \! - \! 2 \! 
\left( \! 1 \! - \! \frac{\eta}{\chi} \! \right)
s \Omega_z^{2} \omega_{A0}^{2} \sin 2 \alpha   
\nonumber \\
\left.
+ \left[ \left( 1 \! - \! \frac{\eta}{\chi} \right) \! 
s \Omega_{z}^{2} - \frac{\eta}{\chi} L_{\varphi} \right]^{2}
+ 2 \frac{\eta}{\chi} L_{\varphi}
\omega_{A0}^{2} \sin 2 \alpha \right\}^{1/2} \!\!< 0. 
\end{eqnarray} 
The two conditions for instability follow straightforward from the above 
expression:
\begin{eqnarray}
s \Omega_{s}^{2} + \frac{\eta}{\chi} \omega_0^2 + \omega_{A0}^{2} < 0~; \\
4 G_{z} \frac{\eta}{\chi} [ s(C_{z} \Omega_{s}^{2} \! - \! C_{s} 
\Omega_{z}^{2}) +
\omega_{A0}^{2} ( C_{z} \cos 2 \alpha  \! + \! C_{s} \sin 2 \alpha)] 
\nonumber \\
-4\sin^{2} \alpha \omega_{A0}^{2} \left( \frac{\eta}{\chi} \omega_{0}^{2} 
\! + \! s \Omega_{s}^{2} \right) \! - \! 2 \! \left( \! 1 \! - \! 
\frac{\eta}{\chi} \! \right)
s \Omega_z^{2} \omega_{A0}^{2} \sin 2 \alpha   
\nonumber \\
+ \left[ \left( 1 \! - \! \frac{\eta}{\chi} \right) \! 
s \Omega_{z}^{2} - \frac{\eta}{\chi} L_{\varphi} \right]^{2}
+ 2 \frac{\eta}{\chi} L_{\varphi}
\omega_{A0}^{2} \sin 2 \alpha > 0. 
\end{eqnarray}
These two conditions look like the Solberg--H{\o}iland conditions
(Tassoul 2000), but with additional terms due to the magnetic field
and diffusive effects. Equations (26) and (27) are equivalent to
the condition $a_{0} < 0$ with the accuracy in terms $\sim (\lambda_{c}/
\lambda)^{2}$. Therefore, the boundary between the stable and unstable
regions can be determined by these conditions with the same accuracy. 
If $B=0$ and $\eta=0$, equation (27) reduces 
to the well-known Goldreich-Schubert-Fricke criterion of instability, 
stating that differential rotation is always unstable if $\Omega_{z}^{2} 
\neq 0$ (Goldreich \& Schubert 1967, Fricke 1968). In the case of a 
non-rotating star ($\Omega=0$) with $B_{\varphi} \gg B_{p}$ when the 
terms containing $\omega_{A0}$ can be neglected, equation (27) yields 
$\eta^{2} L_{\varphi}^{2}/ \chi^{2} > 0$ that is always satisfied. 
Therefore, the star with a presumably toroidal field is always 
unstable that was first argued by Tayler (1973). Generally, if the 
poloidal and toroidal fields are more or less comparable we have 
$L_{\varphi} \sim \omega_{A0}^{2} (kH)^{-2} \ll \omega_{A0}^{2}$, 
and the terms proportional $L_{\varphi}$ can be neglected.    

\subsection{Instability in a weak magnetic field}

We consider equations (26) and (27) in the limiting cases of weak 
and relatively strong magnetic fields. We refer the magnetic field 
weak if $\Omega^{2} \gg \omega_{A0}^{2}$ (it is assumed that $s 
\Omega_{s}^{2} \sim s \Omega_{z}^{2} \sim \Omega^{2}$). This condition 
can be written as
\begin{equation}
B_{p} < B_{cr} = \sqrt{4 \pi \rho} \Omega/k \approx 6 \times 10^{2} 
\sqrt{\rho} \lambda_{8} \Omega_{-5} 
\;\; {\rm G}
\end{equation}   
where $\lambda_{8} = \lambda/10^{8}$ cm, $\Omega_{-5}= \Omega /
10^{-5}$ s, and $\rho$ is measured in g/cc. The quantity $B_{cr}$
depends on $\rho$ and, in general, the same field can be 
considered weak in the inner layers and strong in the outer layers. 
Note that the field should also satisfy the condition $c_{\rm s} > 
c_{\rm A}$ (equation (6)) for validity of our analysis. It can be 
convenient to represent the inequality (28) in terms of the equatorial 
velocity, $V = \Omega R$, where $R$ is the stellar radius. Then, we 
have from equation (28)
\begin{equation}
V > 17 B_{p3} R_{11} \lambda_{8}^{-1} \rho^{-1/2} \;\; {\rm km/s},
\end{equation} 
where $B_{p3}= B_{p}/10^{3}$G, $R_{11}=R/10^{11}$cm. 

In a weak field, equations (26)-(27) yield
\begin{eqnarray}
s \Omega_{s}^{2} + \frac{\eta}{\chi} \omega_0^2  < 0~; \\
4 G_{z} \frac{\eta}{\chi}  s(C_{z} \Omega_{s}^{2} \! - \! C_{s} 
\Omega_{z}^{2}) + \left( 1 \! - \! \frac{\eta}{\chi} 
\right)^{2} \! (s \Omega_{z}^{2})^{2} > 0. 
\end{eqnarray}
The criteria turn out to be independent of the magnetic field
in this limit and, under certain conditions, can be satisfied.

Since $\omega_{0}^{2} > 0$ in the radiative zone, the inequality
(30) can be satisfied only if $\Omega$ decreases with
the cylindrical radius. Taking into account that, typically, 
$\eta/\chi \sim 10^{-4}$ and $\omega_{0}^{2} \sim 10^{-6} - 
10^{-7}$ s$^{-2}$ in the radiative zone and assuming that $|s 
\Omega_{s}^{2}| \sim \Omega^{2}$, we obtain that the condition 
$\eta \omega_{0}^{2} /\chi < |s \Omega_{s}^{2}|$ is satisfied 
if $\Omega > 10^{-5} \omega_{03}$ s$^{-1}$ where $\omega_{03}= 
\omega_{0}/10^{-3}$s. Therefore, the instability caused by 
the condition (30) may occur in relatively rapidly rotating stars 
with $\Omega > 10^{-5} \omega_{03}$ s$^{-1}$ or $V > 10 R_{11} 
\omega_{03}$ km/s if $\partial \Omega / \partial s < 0$.  

The condition (31) depends on the vertical gradient 
of the angular velocity. If $\Omega_{z}^{2} = 0$ then equation
(31) yields 
\begin{equation}
4 s \frac{\eta}{\chi} \; C_{z} G_{z} \Omega_{s}^{2} > 0.
\end{equation}
We have $C_{z} G_{z}< 0$ since both $\vec{C}$ and $\vec{G}$ are 
approximately radial and oppositely directed in the radiative
zone if the star is far from the rotational distortion. Therefore,
the inequality (32) is equivalent to $\Omega_{s}^{2} < 0$, and
both slowly and rapidly rotating stars can be unstable in 
this case if the wavelength of perturbations satisfies the 
condition (17).

If $\Omega_{z}^{2} \neq 0$ then the vertical dependence of the
angular velocity can produce either stabilizing or destabilizing
effect depending on $\Omega(z)$. If rotation is slow and $\eta 
\omega_{0}^{2} / \chi > \Omega^{2}$, then the criterion (31) is 
equivalent to
\begin{equation}
G_{z}(C_{z} \Omega_{s}^{2} - C_{s} \Omega_{z}^{2}) > 0.
\end{equation} 
In slowly rotating stars, $\vec{G}$ and $\vec{C}$ are 
approximately radial and directed inward and outward, 
respectivetely. Then, we have
\begin{equation}
G_{z} \approx - G \cos \theta, \;\; C_{s} \approx C \sin
\theta, \;\; C_{z} \approx C \cos \theta,
\end{equation}
where $\theta$ is the polar angle. Substituting these 
expressions into equation (33), we obtain 
\begin{equation}
\Omega_{s}^{2} \cos^{2} \theta - \Omega_{z}^{2} \sin \theta 
\; \cos \theta < 0. 
\end{equation}
Hence, $\Omega_{z}^{2}$ provides a destabilizing influence 
if $\Omega$ increases with the distance from the equator and a 
stabilizing influence in the opposite case. If $\Omega_{s}^{2} 
< 0$ and $\Omega_{z}^{2} < 0$ then instability occurs only in a 
cone around the rotational axis, and the opening angle of this 
cone depends on the radial and vertical lengthscale of $\Omega$. 
On the contrary, if $\Omega_{s}^{2} > 0$ and $\Omega_{z}^{2} > 
0$ then the instability arises only in a region around the 
equatorial plane. 

Rapidly rotating stars with a sufficiently strong vertical 
differential rotation, $|s \Omega_{z}^{2}| > \eta \omega_{0}^{2} 
/ \chi$, can be unstable independently of the particular 
shape of $\Omega(z)$ if $\lambda$ satisfies condition (17), and 
the instability occurs everywhere within the radiative zone.

\subsection{Instability in a strong magnetic field}

We refer the magnetic field strong if 
\begin{equation}
B_{p} > B_{cr} = 6 \times 10^{2} \sqrt{\rho} \lambda_{8} \Omega_{-5} 
\;\; {\rm G}
\end{equation}
(but the condition (6) still holds) because such a field can provide 
a strong influence on the stability properties. Equation (36) can be
rewritten as 
\begin{equation}
V < 17 B_{p3} R_{11} \lambda_{8}^{-1} \rho^{-1/2} \; {\rm km/s}.
\end{equation}
Under this condition, the criteria (26) and (27) transform into
\begin{eqnarray}
\frac{\eta}{\chi} \omega_0^2 + \omega_{A0}^{2} < 0~; \\
2 G_{z} \frac{\eta}{\chi} 
( C_{z} \cos 2 \alpha  \! + \! C_{s} \sin 2 \alpha)  
- 2 \sin^{2} \alpha \left( \frac{\eta}{\chi} \omega_{0}^{2} 
+ s \Omega_{s}^{2} \right) 
\nonumber \\
- \left( \! 1 \! - \! 
\frac{\eta}{\chi} \! \right) s \Omega_z^{2} \sin 2 \alpha  > 0. 
\end{eqnarray}
The first criterion never applies in the radiative zone since
$\omega_{0}^{2} > 0$, and we consider only the condition (39). 
Note that this criterion turns out to be independent of the 
strength of the magnetic field but depends on its direction.

In slowly rotating stars with $\eta \omega_{0}^{2} / \chi >
\Omega^{2}$, we have from equation (39) 
\begin{equation}
G_{z}  
( C_{z} \cos 2 \alpha  \! + \! C_{s} \sin 2 \alpha)  
- \omega_{0}^{2} \sin^{2} \alpha  > 0
\end{equation}
or, using equations (34),
\begin{equation}
(\cos \theta \cos \alpha - \sin \theta \sin \alpha)^{2}< 0
\end{equation}
that never applies. Therefore, the radiative zones of slowly
rotating stars with a strong magnetic field satisfying the
inequality (36) should be secularly stable. 

Consider the case of a relatively rapid rotation, $\Omega^{2} 
> \eta \omega_{0}^{2}/ \chi$, or $V> 10 R_{11} \omega_{03}$ 
km/s. Combining this condition with the inequality (37), we 
obtain the range of $V$ where rotation can be considered rapid 
and the magnetic field strong
\begin{equation}
17 B_{p3} R_{11} \lambda_{8}^{-1} \rho^{-1/2} \; {\rm km/s} > V > 
10 R_{11} \omega_{03} \; {\rm km/s}. 
\end{equation}
Equation (39) yields within this range
\begin{equation}
\sin \alpha (\Omega_{s}^{2} \sin \alpha + \Omega_{z}^{2} \cos 
\alpha) < 0
\end{equation}
(we neglect the term $\eta/ \chi$ compared to unit). If 
rotation is cylindrical and $\Omega_{z}^{2}=0$ then instability
occurs if the angular velocity decreases with the cylindrical
radius. 

Stability properties can be more complicated, however, if  
rotation departs from a simple cylindrical law. Consider, for 
example, the so-called shellular rotation when $\Omega$ is
approximately constant on spherical shells, $\Omega= \Omega(r)$,
$r$ is the spherical radius. Such rotation can be a reasonable 
approximation, for example, in the radiative zones 
of massive main sequence stars (see, e.g., Zahn 1992, Urpin, 
Shalybkov \& Spruit 1996). For a shellular rotation, we have
$\Omega_{s}^{2} = \Omega_{r}^{2} \sin \theta$ and $\Omega_{z}^{2} 
= \Omega_{r}^{2} \cos \theta$ where $\Omega_{r}^{2} = \partial 
\Omega^{2}/ \partial r$. Then, the condition (43) reads
\begin{equation}
\Omega_{r}^{2} \sin \alpha ( \sin \alpha \sin \theta +
\cos \alpha \cos \theta) < 0.
\end{equation}  
If the angular velocity decreases with the spherical radius,
$\Omega_{r}^{2} < 0$, the instability occurs in the region 
where
\begin{equation}
\sin \alpha \cos (\alpha - \theta) > 0.
\end{equation}
There is no generally accepted point of view regarding the 
geometry and strength of the magnetic field inside stars.
For the purpose of illustration, we consider therefore the 
simplest magnetic configuration that can exist in the radiative 
zone
\begin{equation}
B_{r} = f(r) (1 - 3 \cos^{2} \theta), \;\;\; B_{\theta} =
F(r) \sin \theta \cos \theta,
\end{equation}
where $f$ and $F$ are functions of the spherical radius alone
and satisfies the divergence-free condition
\begin{equation}
F = \frac{1}{r} \frac{d(r^{2} f)}{dr}.
\end{equation}
The magnetic field (46) has the same geometry as the velocity 
pattern of meridional circulation caused by shellular rotation 
with $\Omega = \Omega(r)$ (Zahn 1992, Urpin, Shalybkov \& Spruit 
1996, Talon et al. 1997)). Therefore, the assumption (46) is a 
good approximation if the field is frozen-in a slow circulation 
flow. Generally, the field in radiative zones can have a more 
complex geometry but the simple model (46) allows to understand 
qualitatively how the magnetic field influences the region of 
instability.  

We have for the magnetic field (46) 
\begin{eqnarray}
\sin \alpha = \frac{\vec{e}_s \cdot \vec{B}}{B_{p}} =
\frac{\sin \theta}{B_p}[f (1 - 3 \cos^{2} \theta) + F \cos^{2}
\theta ],     \nonumber \\
\cos \alpha = \frac{\vec{e}_z \cdot \vec{B}}{B_{p}} =
\frac{\cos \theta}{B_p}[f (1 - 3 \cos^{2} \theta) - F \sin^{2}
\theta ].
\end{eqnarray}
Substituting these expressions into equation (44), we obtain 
\begin{equation}
(1 - 3 \cos^{2} \theta) [ 1 - (1 - q) \cos^{2} \theta] > 0,
\end{equation}
where $q= d \ln f / d \ln r$. The criterion (49) is fulfilled at 
$\cos \theta \approx 0$, and the region near the equator is secularly 
unstable. Near the rotation axis where $\cos \theta \approx 1$, 
equation (49) yields 
\begin{equation}
\frac{df}{dr} < 0,
\end{equation} 
and this region can be unstable only if
the radial field component decreases to the surface. The regions 
of instability as a function of the parameter $q$ are shown in 
Table 1.  

\begin{table*}
\begin{center}
Table 1. \\
The dependence of the region of instability on $q$ for a rapid shellular
rotation. \\
\end{center}
\begin{center}
\begin{tabular}{|l|l|l|l|}
\hline
$q$    &  $q > 0$ &  $0 > q > -2$ &  $-2 > q$      \\
\hline
       &          &               &                 \\
Unstable  &  $\cos \theta < \sqrt{3}/3$ &  $\cos \theta < \sqrt{3}/3 $  &
             $\cos \theta < 1/\sqrt{1-q}$   \\
region    &   &  $\cos \theta > 1/\sqrt{1-q}$ &
             $\cos \theta > \sqrt{3}/3$ \\
       &          &               &                \\
\hline
\end{tabular}
\end{center}
\end{table*}

There is a principal difference between the cases $q < 0$ and
$q > 0$. The first case can correspond to the field generated 
somewhere in the central region of the star, for example, in 
the convective core that exists in massive stars. The second 
case is likely more relevant to the field generated in the 
surface layers. For instance, this can be typical for the
late-type stars which have outer convective zones where 
turbulent dynamo operates. If the field increases with $r$ and 
$q>0$ then the instability occurs only in the region 
with openning angle $\approx 36^{\circ}$ around the equator.  
A wide region around the rotation axis with $\theta < 54^{\circ}$
is stable. On the contrary, if the field decreases to the
surface and $q<0$ then there are two different unstable regions. 
For a slowly decreasing field, $0>q>-2$, the instability occurs 
in a relatively wide unstable region around the equator (with 
the openning angle $\theta > 54^{\circ}$) and in a narrow region 
around the axis (with the openning angle $\arccos 1/\sqrt{1-q}$). 
If the field decreases rapidly with $r$ and $q < -2$, then the 
instability occurs in a region with $\theta < 36^{\circ}$ around 
the rotation axis and in a narrow region around the equator 
(with the openning angle $\arccos 1/\sqrt{1-q}$). Note that this 
openning angle can be very small if the field decreases rapidly.

\section{The growth rate of instability}

Consider the growth rate of the modes assuming that the dissipative 
frequences $\omega_{\eta}$, $\omega_{\nu}$ and $\omega_{\chi}$ are 
smaller than the dynamical frequences $\omega_{A}$, $\omega_{g}$ and 
$\Omega$ because this case is of particular interest for stellar
radiative zones. Under this assumption, the dispersion equation (15) 
can be solved by making use of the perturbation procedure. We expand 
$\gamma$ as $\gamma = \gamma^{(0)} + \gamma^{(1)} + ...$ where 
$\gamma^{(0)}$ and $\gamma^{(1)}$ are terms of the zeroth and first
order in dissipative frequencies, respectively. The corresponding 
expansion should be made for the coefficients of equation (15) as 
well. In the zeroth order when dissipation is neglected, equation 
(15) reduces to a quadratic equation,
\begin{equation}
\gamma^{4} + (2 \omega_{A}^{2} + \omega_{g}^{2} + Q^{2}) 
\gamma^{2} + \omega_{A}^{2} ( \omega_{A}^{2} + \omega_{g}^{2}
+ F^{2})= 0, 
\end{equation}
This equation describes four modes with the frequencies
\begin{eqnarray}
\gamma^{(0) \; 2}_{1,2} = \!- \!\omega_{A}^{2}\! -\! \frac{\omega_{g}^{2} 
+Q^{2}}{2} + \! \sqrt{\frac{(\omega_{g}^{2} + Q^{2})^{2}}{4}
+ \! 4 \omega_{A}^{2} \Omega^{2} \frac{k_z^{2}}{k^{2}}},  \\
\gamma^{(0) \; 2}_{3,4} =\! - \!\omega_{A}^{2}\! -\! \frac{\omega_{g}^{2} 
+Q^{2}}{2} - \! \sqrt{\frac{(\omega_{g}^{2} + Q^{2})^{2}}{4}
+ \! 4 \omega_{A}^{2} \Omega^{2} \frac{k_z^{2}}{k^{2}}}.
\end{eqnarray}
The fifth root of equation (15) is vanishing in the zeroth 
approximation, $\gamma_{5}^{(0)}= 0$. Equation (51) has unstable 
solutions with Re $\gamma > 0$ if one of the conditions
\begin{equation}
2 \omega_{A}^{2} + \omega_{g}^{2} + Q^{2} < 0, \;\; \omega_{A}^{2} 
+ \omega_{g}^{2} + F^{2} < 0, 
\end{equation}
is satisfied. The first of these inequalities describes the 
well-known Rayleigh instability, and the second condition
corresponds to the magnetorotational instability (Velikhov
1959, Chandrasekhar 1960, Balbus \& Hawley 1991). Most likely, 
neither of the 
inequalities (54) is fulfilled in radiative zones because of 
a strong stabilizing effect of stratification. The only 
exception are pertutbations with $\vec{k} \! \parallel \! 
\vec{G}$ for which $\omega_{g}^{2} = 0$ (Urpin 1996). Note that, 
the stabilizing effect of stratification is much suppressed 
for the instability considered in this paper (see equation (18)) 
compared to the Rayleigh and magnetorotational instabilities, 

The fifth root of equation (15) describes a secular mode that is 
linear in dissipative effects. To calculate this root we should 
keep in equation (15) only the terms linear in dissipative 
frequencies assuming that $\gamma_{5}$ is linear as well. Then, 
the fifth root is approximately given by
\begin{equation}
\gamma_{5}^{(1)} \approx - \omega_{\chi} \; \frac{F^{2} + 
\omega_{A}^{2} + (\eta/\chi) \omega_{g}^{2}}{F^{2} + 
\omega_{A}^{2} + \omega_{g}^{2}}.
\end{equation}
The order of magnitude estimate of this root reads
\begin{equation}
\gamma_{5}^{(1)} \sim \omega_{\chi} \; \frac{\Omega^{2}}{\omega_{g}^{2}}
\sim 4 \times 10^{-10} \chi_{7} \omega_{03}^{-2} R_{11}^{-2}
\lambda_{8}^{-2} V_{7}^{2} \; {\rm s}^{-1},
\end{equation}
where $\chi_{7} = \chi/10^{7}$ cm$^{2}$/s and $V_{7} = V / 10^{7}$
cm/s. Then, the growth time of instability is
\begin{equation}
\tau = 1/ \gamma_{5}^{(1)} \sim 10^{2} \chi_{7}^{-1} \omega_{03}^{2} 
R_{11}^{2} \lambda_{8}^{2} V_{7}^{-2} \; {\rm yrs},
\end{equation}
The growth time can be rather short if the star rotates rapidly, 
$V_{7} \sim 1$.
 
\section{Discussion}

We have considered the diffusive instability of differentially
rotating magnetized stellar radiative zones. The instability 
considered is a generalization of the well-known 
Goldreich-Schubert-Fricke instability for the case of magnetic 
stars. In non-magnetic stars, this instability occurs if the 
angular velocity depends on the vertical coordinate, $\Omega = 
\Omega(z)$. The magnetic field, however, changes drastically 
the stability properties of differentially rotating stars. Even 
in a weak field satisfying the inequality (28), the magnetic 
analogue of the Goldreich-Schubert-Fricke instability may occur 
in a region where $\Omega$ depends on the cylindrical radius alone, 
$\Omega =\Omega(s)$. The necessary condition of instability in this 
case is a decrease of the angular velocity with the cylindrical 
radius. If $\partial \Omega/ \partial s < 0$ than the instability
may occur in both slowly and rapidly rotating stars. If rotation 
is more complicated, $\Omega=\Omega(s, z)$, then the vertical 
dependence of $\Omega$ can provide either stabilizing or 
destabilizing effect depending on the value and particular 
shape of $\Omega(s, z)$. In rapidly rotating stars with the
equatorial velocity $V > 10 R_{11} \omega_{03}$ km/s, the vertical 
dependence of $\Omega$ plays a destabilizing role, and the 
radiative zones of such stars should be unstable. In slowly 
rotating stars, the vertical dependence destabilizes differential
rotation if $\partial \Omega /\partial z > 0$ and stabilizes if  
$\partial \Omega /\partial z < 0$. In such stars, generally, the 
instability can occur only in some regions near the rotation axis 
or equator depending on the particular shape of $\Omega(s, z)$.  
  
In a strong magnetic field with $B_p > B_{cr}$ (but still 
satisfying the condition $c_{\rm s} > c_{\rm A})$, the 
stability properties are essentially different. The secular
instability does not occur in slowly rotating stars with the 
equatorial velocity $V < 10 R_{11} \omega_{03}$ km/s. 
However, if rotation is rapid and satisfies 
the condition (42) then the radiative zone or its fraction can be 
subject to instability. The geometry of unstable regions is
sensitive to the particular shape of $\Omega(s, z)$. Even in our 
very simplified model where the axisymmetric magnetic field is 
considered, the instability may occur not in the whole radiative 
zone but only within some regions restricted by cones in the 
meridional plane. For example, in the case of a shellular rotation 
when $\Omega$ depends on the spherical radius alone, the polar 
angle corresponding to the unstable regions is determined by the radial 
dependence of the magnetic field and is shown in Table 1. If the 
radial magnetic field increases to the surface then the instability 
arises only in the region around the equator. In the opposite
case when $B_{r}$ decreases to the surface, the instability
occurs both around the equator and rotation axis. The 
openning angle of unstable regions depends on how fast the
magnetic field decreases with $r$. 
       
The growth time of instabilities is relatively short and, 
likely, it operates in a non-linear regime. We can estimate 
the saturation velocity using the mixing-length model (e.g., 
Schwarzschild 1958) that assumes that the turn-over time of 
turbulence generated by instability is of the order of the 
growth time of this instability. In a vanishing magnetic field,
this simple approach provides a quantitatively correct estimate 
of the saturation velocity generated by the 
Goldreich-Schubert-Fricke instability in stellar conditions 
(Korycansky 1991) and accretion disks (see Urpin 2003c and Arlt 
\& Urpin 2004). Then, the saturation velocity in a turbulent cell 
with the lengthscale $\lambda$ can be estimated as $v_{T} (\lambda) 
\sim \lambda \; \gamma(\lambda)$ or, using equation (57), as
\begin{equation}
v_{T} \sim 0.04 \chi_{7} \omega_{03}^{-2} R_{11}^{-2}
\lambda_{8}^{-1} V_{7}^{2} \;\; {\rm cm/s}.
\end{equation} 
It is worth remarking that $v_{T}$ is larger for small scale 
motions. 

Turbulent motions can enhance transport and mix the material 
in the unstable regions. The coefficient of turbulent 
diffusion, $\nu_{T}$, can be estimated as a product of the 
turbulent velocity and lengthscale. Then, we have 
\begin{equation}
\nu_{T} \sim v_{T} \lambda \sim 4 \times 10^{6} \chi_{7} 
\omega_{03}^{-2} R_{11}^{-2} V_{7}^{2} \;\; {\rm cm^{2}/s}.
\end{equation}
Turbulent diffusion caused by the instability is sufficient to mix 
the unstable zone on a relatively short timescale. Using equation
(59), we can estimate the timescale of mixing as
\begin{equation}
\tau_{mix} \sim \frac{R^{2}}{\nu_{T}} \sim
6 \times 10^{7} \chi_{7}^{-1} \omega_{03}^{2} R_{11}^{4} V_{7}^{-2}
\;\; {\rm yrs}
\end{equation} 
that can be substantially shorter than the lifetime of many stars.
Therefore, elements from a deep interior can likely be transported 
to the stellar surface by turbulence within the unstable zones. 

This conclusion is of particular importance for strongly magnetized 
stars with the rotation velocity satisfying the condition (42). In 
such stars, the instability leads to a formation of non-spherical 
``turbulent zones'' mixing the material of deep interiors and the 
surface layer. In our very idealized model with 
the axisymmetric magnetic field and shellular rotation, the 
instability can form two ``mixing zones'': the cone turbulent 
region around the rotation axis and the turbulent belt around the 
equator. If rotation or the magnetic field are more complex then 
the geometry of turbulent ``mixing zones'' can be more complicated. 
For example, if the magnetic field is non-axisymmetric then the 
distribution of unstable regions should be non-axisymmetric as well. 
The material should be well mixed in the ``mixing zones'' and, 
likely, this mixed material can reach even the atmosphere if the 
unstable region extends to the surface. If the magnetic field is 
axisymmetric then the pecular regions, where the ``mixing zones'' 
reach the surface, have likely the belt-like structure or form the 
polar spot with a pecular composition. If the field is non-axisymmetric
then the instability can lead to a formation of the spot-like 
structure with chemical pecularities.

{}

\end{document}